\documentclass[twocolumn,letterpaper,aps,prc,superscriptaddress,nofootinbib,showpacs,floatfix]{revtex4-1}
\usepackage{comment}
\usepackage{setspace}
\usepackage{color}
\usepackage{indentfirst}
\usepackage{xspace}
\usepackage{verbatim}
\usepackage{epstopdf}
\usepackage[export]{adjustbox}
\usepackage[T1]{fontenc}
\usepackage{float}
\usepackage{wrapfig}
\usepackage{graphicx}
\usepackage{subcaption}
\usepackage{diffcoeff}
\usepackage{amsmath}
\usepackage{amssymb}
\usepackage{amsthm}
\usepackage{array}
\usepackage{tabularx}
\usepackage[utf8]{inputenc}
\graphicspath{ {figures/} }
\usepackage{lineno}
{\large }

\usepackage{natbib}
\setcitestyle{square, comma, numbers,sort \& compress, super}
\usepackage{appendix}
\usepackage{dcolumn}
\usepackage{bm}
\usepackage[colorlinks=true, linktocpage=true, linkcolor=blue, citecolor=blue]{hyperref}
\usepackage[a4paper]{geometry}
\begin{document}
	\title{Multiplicity and  Transverse Spherocity dependence of  $\langle p_{\rm T} \rangle$ fluctuations of charged particles  in p$-$p collisions at $\sqrt{s}$ =  7 and 13 TeV}
	\author{ Subhadeep~Roy }
	\email{subhadeep.roy@cern.ch}
	\affiliation{Indian Institute of Technology Bombay, Mumbai 400076, India}
	
	\author{ Tulika~Tripathy }
	\email{tulika.tripathy@cern.ch}
	\affiliation{Indian Institute of Technology Bombay, Mumbai 400076, India}
	
	\author{ Sadhana~Dash }
	\email{sadhana@phy.iitb.ac.in}
	\affiliation{Indian Institute of Technology Bombay, Mumbai 400076, India}
	
	\begin{abstract}
		
		The multiplicity dependence of event-by-event fluctuations in mean transverse momentum, $\langle p_{\rm T} \rangle$,  of charged particles has been studied in p$-$p collisions at  $\sqrt{s}$ = 7 TeV and 13 TeV  using the 
		PYTHIA 8 event generator. The charged particles were selected in  kinematic range of  $0.15 < p_{\rm T}<2$  GeV/\textit{c} and $|\eta| < 0.8$.  The dynamical fluctuations  would indicate towards the correlated emission of particles. The measurements in A$-$A and p$-$p collisions has shown a decrease in the strength of $ \langle p_{\rm T} \rangle$ fluctuations with the average charged particle multiplicity. The effects of various microscopic processes like color reconnection and multi-partonic interactions has been studied. A minimal dependency on the collision energy is also observed. 
		Furthermore, the fluctuation observables are investigated in the intervals of transverse spherocity in order to comprehend the relative contributions resulting from hard scattering and underlying events. 
		The present  study would act as a baseline for future measurements in A$-$A as well as p$-$p collisions at the LHC.
		
	\end{abstract}
	
	\maketitle
	

	\section{Introduction}
	The search for event-by-event fluctuations in dynamical quantities, such as mean  transverse momentum ($ \langle p_{\rm T} \rangle$)  in  heavy-ion collisions  is primarily motivated by the search for the presence  and characterization  of the phase transition between a quark-gluon plasma (QGP) and hadron gas state  \cite{star2005, alice2014,trainor1}.  One can quantify this search by looking for excess fluctuations of thermodynamic quantities like temperature. The event-wise $\langle p_{\rm T} \rangle$  can be taken as a proxy for local temperature and hence the study of its event-by-event fluctuation has a thermodynamic context. The  temperature fluctuations are 
	related to the heat capacity of the system. The non-monotonic behaviour of heat capacity characterizes a phase transition and hence the study of event-by-event $ \langle p_{\rm T} \rangle$ fluctuations could reveal important information about the formation of QGP \cite{stodolsky}.
	
	The recent  measurements by the STAR \cite{star2005,star2019} and ALICE \cite{alice2014} experiment at  RHIC and LHC reported that the strength of dynamical  $ \langle p_{\rm T} \rangle$ fluctuations  decreased  with event multiplicity (or centrality) in heavy-ion collisions. The observation was attributed to the onset of thermalization, collectivity, jet suppression, presence of di-jets and minijets and other processes.  The interesting studies showing
	the $ \langle p_{\rm T} \rangle$ fluctuations and their correlations with radial flow and elliptic flow in heavy-ion collisions can be found in the references \cite{mptvoloshin,schenke}. It remains questionable whether the system created in  small systems like  p$-$p or p$-$A collisions  exhibit collectivity as observed in A$-$A collisions. This can be addressed  by studying particle production mechanisms, their correlations etc. as a function of the event shape and particle multiplicity. 
	Recent experimental results of $ \langle p_{\rm T} \rangle$ in p$-$p collisions \cite{alice2014} suggest that, high-multiplicity events are mostly produced by multiple parton interactions (MPIs) \cite{mpi}. 
	The data exhibited a power-law trend  with charged particle multiplicity density, as expected from the independent superposition scenario of independent multi-partonic interactions. However, the slight deviation of the 
	power-law index from  $0.5$ indicated towards non-trivial fluctuations. In PYTHIA 8 \cite{pythia8}, the observed strong correlation  between the particle multiplicity and $\langle p_{\rm T} \rangle$  has been attributed to the mechanism of color reconnections (CR) between hadronizing strings \cite{color0,color1,color2}.  The color strings can also overlap with each other to form color ropes that act coherently in high-multiplicity events \cite{rope1,rope2}. This overlap region causes a pressure gradient, thereby shoving the strings in the outward direction creating similar effects as observed in hydrodynamic scenario. The color reconnections together with the rope hadronization could explain some of the high-multiplicity observables in p$-$p collisions\cite{sdashrope1,sdashrope2,sdashrope3}. A similar  mechanism of collective relativistic string hadronization is also implemented in the EPOS model which describes a wealth of LHC data in p$-$p, p$-$Pb, and Pb$-$Pb collisions \cite{epos}.
	\par
	In this work, an attempt has been made to study the $ \langle p_{\rm T} \rangle$ fluctuations  in p$-$p collisions at collision energies, $\sqrt{s}$ = 7 TeV and 13 TeV as a function of multiplicity and different tansverse spherocity classes using PYTHIA 8 generator. 
	The fluctuation is generally quantified  by the the second moment of the distribution and the two-particle correlator has been used to study the interplay of various processes. The higher moments, namely the third and fourth moment were also studied as a function of multiplicity to shed additional lights on higher order correlations in multi-particle production mechanism.  The additional effects of color reconnection as well as the formation of ropes were also studied. This study would act as a baseline for future measurement in A$-$A and p$-$p systems.
	\section{Transverse Spherocity}
\begin{figure}[h!]
		\centering
		\includegraphics[width=0.7\linewidth]{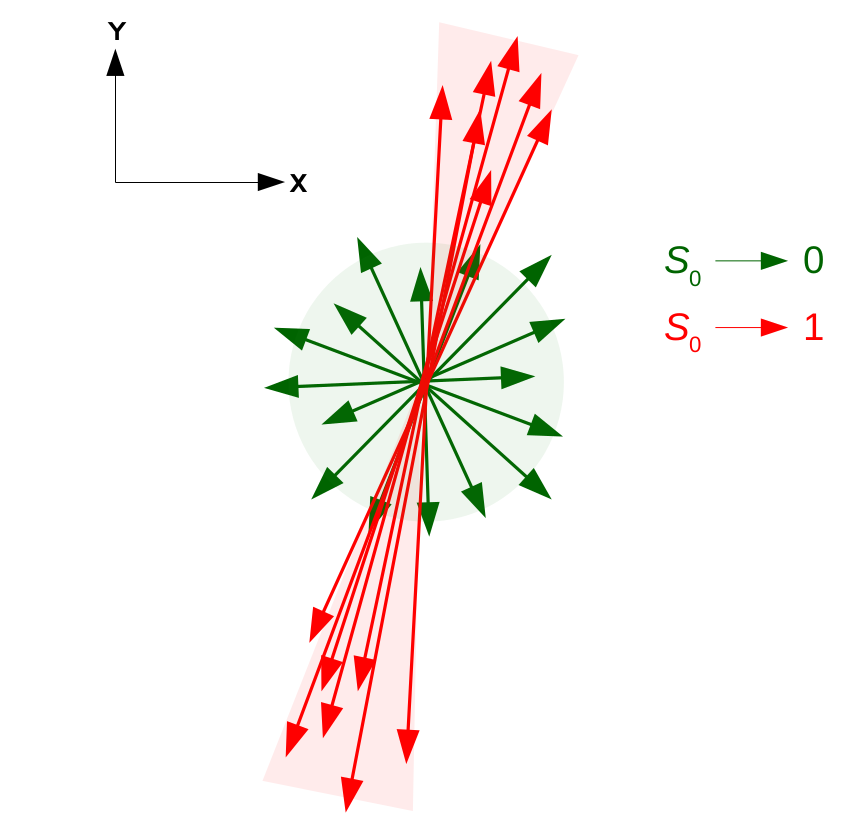}
		\caption{Illustration of the typical topologies for the two $S_{0}$ limits in the azimuthal plane.}
		\label{spherocity_dia}
	\end{figure}

	The study of various observables in terms of event shape variables such as transverse spherocity, thurst etc. helps to disentangle the contributions from hard and soft (scattering) processes of particle production. 
	The transverse spherocity is a tool which characterises the event into jet-like or isotropic events \cite{sphero1,sphero2,sphero3} . 
	It is defined  for a unit transverse vector , ${\hat{n}(n_{\rm T},0)}$, which minimizes the ratio 
	\begin{equation}
	S_{0} = \frac{\pi^{2}}{4}\underset{\hat{ n}}{\rm min} \left ({\frac{\sum\limits_{i} |\overrightarrow{p_{\rm T_{i}}} \times \hat{ n}|}{\sum\limits_{i} p_{\rm T_{i}}}}\right )^{2} 
	\end{equation} 
	This unit vector coincides with the transverse direction of the momentum, $\overrightarrow{p_{\rm T}}$ which makes the $S_{0}$ infrared  and  collinear safe. The sum runs over all the particles considered in the kinematic acceptance. Events with $S_{\rm 0} \rightarrow 0$ are likely to be dominated by a single hard scattering, producing a jet-like structure ({\bf Jetty}). In contrary, $S_{\rm 0} \rightarrow 1$ implies that particles are produced through several softer interactions, giving an {\bf isotropic} distribution. An illustration of the two limits in the azimuthal plane is presented in Figure \ref{spherocity_dia}.
	To replicate the same circumstances as the ALICE experiment at the LHC, the transverse spherocity distributions are chosen in the pseudo-rapidity range of $|\eta|<$ 0.8 with a minimum constraint of $5$ charged particles with $p_{\rm T}>$  0.15 GeV/{\textit{c}}.
	The various percentiles of the whole $S_{0}$ distribution are used for the selection of Jetty (0-20\%) and isotropic (80-100\%) event classes. Transverse spherocity will henceforth be referred to as \textit{spherocity} for the sake of simplicity.
	\section*{Observables}
	The mean value of the transverse momentum, obtained for a particular multiplicity class can be defined as 
	\begin{equation}	
	\left\langle \!\left\langle p_{\rm T}\right\rangle\!\right\rangle=\left\langle\frac{\sum_{i=1}^{N_{\rm ch}} p_{{\rm i}}}{N_{\rm ch}} \right\rangle
	\end{equation}		
	where $N_{\rm ch}$ denotes the number of charged particles in a single event and $p_{{\rm i}}$ represents the transverse momentum of the $i^{\rm th}$ particle in the event. The first average is determined over all of the $p_{{\rm i}}$ in an event, while the second average is carried out over all of the events included in a certain multiplicity class. From this point on, the mean transverse momentum of the particles will be denoted by $\left\langle \!\left\langle p_{\rm T}\right\rangle\!\right\rangle$.\par 
	The two-particle transverse momentum correlator, $\left\langle \Delta p_{\rm i}\Delta p_{\rm j}\right\rangle$, is constructed as $-$ 
	\begin{widetext}
	\begin{equation}
	\left\langle \Delta p_{\rm i}\Delta p_{\rm j}\right\rangle=\left\langle\frac{\sum_{i, j \ne i}(p_{\rm i}-\left\langle \left\langle p_{\rm T}\right\rangle\right\rangle ) (p_{\rm j}-\left\langle \left\langle p_{\rm T}\right\rangle\right\rangle )}{N_{\rm ch}(N_{\rm ch}-1)} \right\rangle		
	\end{equation}
	Similarly, the three-particle and four-particle correlators are given by 
	\begin{equation}
	\left\langle \Delta p_{\rm i}\Delta p_{\rm j}\Delta p_{\rm k}\right\rangle=\left\langle\frac{\sum_{i, j \ne i, k \ne i, j}(p_{\rm i}-\left\langle \left\langle p_{\rm T}\right\rangle\right\rangle ) (p_{\rm j}-\left\langle \left\langle p_{\rm T}\right\rangle\right\rangle ) (p_{\rm k}-\left\langle \left\langle p_{\rm T}\right\rangle\right\rangle )}{N_{\rm ch}(N_{\rm ch}-1)(N_{\rm ch}-2)} \right\rangle		
	\end{equation}
	\begin{equation}
	\left\langle \Delta p_{\rm i}\Delta p_{\rm j}\Delta p_{\rm k} \Delta p_{\rm l}\right\rangle=\left\langle\frac{\sum_{i, j \ne i, k \ne i, j, l \ne i,j,k}(p_{\rm i}-\left\langle \left\langle p_{\rm T}\right\rangle\right\rangle ) (p_{\rm j}-\left\langle \left\langle p_{\rm T}\right\rangle\right\rangle ) (p_{\rm k}-\left\langle \left\langle p_{\rm T}\right\rangle\right\rangle )(p_{\rm l}-\left\langle \left\langle p_{\rm T}\right\rangle\right\rangle )}{N_{\rm ch}(N_{\rm ch}-1)(N_{\rm ch}-2)(N_{\rm ch}-3)} \right\rangle		
	\end{equation}
	These correlators can be written in a more simpler form when expressed in terms of raw moments as follows \cite{skewness} :
	\begin{align}
	\nonumber \langle\!\langle p_{\rm T}\rangle\!\rangle&=\left\langle\frac{Q_1}{N_{\rm ch}}\right\rangle, \\
	\nonumber \left\langle \Delta p_{\rm i} \Delta p_{\rm j}\right\rangle&=\left\langle \frac{Q_{1}^2-Q_2}{N_{\rm ch}\left(N_{\rm ch}-1\right)}  \right\rangle
	-\left\langle\frac{Q_1}{N_{\rm ch}}\right\rangle^2, \\
	\nonumber \left\langle \Delta p_{\rm i} \Delta p_{\rm j} \Delta p_{\rm k}\right\rangle&=\left\langle \frac{Q_1^3-3Q_2Q_1+2 Q_3}{N_{\rm ch}\left(N_{\rm ch}-1\right)\left(N_{\rm ch}-2\right)} \right\rangle-3\left\langle \frac{Q_1^2-Q_2}{N_{\rm ch}\left(N_{\rm ch}-1\right)}  \right\rangle\left\langle\frac{Q_1}{N_{\rm ch}}\right\rangle+2\left\langle\frac{Q_1}{N_{\rm ch}}\right\rangle^3.
	\end{align}
	\begin{multline}
	\nonumber \left\langle \Delta p_{\rm i} \Delta p_{\rm j} \Delta p_{\rm k}\Delta p_{\rm l}\right\rangle=\left \langle \frac{Q_{1}^4 -6Q_{4} +8Q_{1}Q_{3} -6Q_{1}^2 Q_{2} +3Q_{2}^2}{N_{\rm ch} (N_{\rm ch} -1) (N_{ch} -2) (N_{ch} -3 )} \right \rangle- 4 \left\langle \frac{Q_1^3-3Q_2Q_1+2 Q_3}{N_{\rm ch}\left(N_{\rm ch}-1\right)\left(N_{\rm ch}-2\right)} \right\rangle \left \langle \frac{Q_{1}}{N_{\rm ch}}\right \rangle \\+6\left\langle \frac{Q_1^2-Q_2}{N_{\rm ch}\left(N_{\rm ch}-1\right)}  \right\rangle\left\langle\frac{Q_1}{N_{\rm ch}}\right\rangle ^2 -3 \left\langle\frac{Q_1}{N_{\rm ch}}\right\rangle^3.
	\end{multline} 
	where,  $Q_{\rm n}$s are the moments of the transverse momentum distributions in an event and is written as
   \end{widetext}
	\begin{equation}
	\label{defqn}
	Q_{\rm n}=\sum_{i=1}^{N_{\rm ch}} (p_{\rm i})^n,
	\end{equation}
	$p_{\rm i}$ denotes the transverse momentum of the particle $i$, and the sum runs over all the charged particles in an event. The different value of $n$ corresponds to different order of the moment. \par 
	In order to make the analysis less sensitive to the range in $p_{\rm i}$, we normalise the parameters  with $\left\langle \!\left\langle p_{\rm T}\right\rangle \! \right\rangle$. Thus, the final expressions for two-particle and four-particle correlator under the study takes the following forms $-$
	\begin{align}
	\left\langle \Delta p_{\rm i}\Delta p_{\rm j}\right\rangle_{final}=\frac{\sqrt{\left\langle \Delta p_{\rm i}\Delta p_{\rm j}\right\rangle}}{\left\langle\! \left\langle p_{\rm T}\right\rangle \! \right\rangle}
	\label{twopart_eq}
	\end{align}
	\begin{align}
	\left\langle \Delta p_{\rm i}\Delta p_{\rm j} \Delta p_{\rm k} \Delta p_{\rm l}\right\rangle_{final}= \frac{(\left\langle \Delta p_{\rm i}\Delta p_{\rm j} \Delta p_{\rm k} \Delta p_{\rm l}\right\rangle)^{1/4}}{\left\langle \! \left\langle p_{\rm T}\right\rangle \!\right\rangle}
	\label{fourpart_eq}
	\end{align}
	Likewise, in order to make the three-particle correlator dimensionless, it is normalised as follows 
	\begin{align}
	\gamma_{p_{\rm T}}=\frac{\left\langle \Delta p_{\rm i}\Delta p_{\rm j} \Delta p_{\rm k}\right\rangle}{{\left\langle \Delta p_{\rm i}\Delta p_{\rm j}\right\rangle}^{3/2}}
	\label{standskew_eq}
	\end{align}
	Now the obtained version of the three-particle correlator is called standardized skewness , and further has a system size or centrality dependence, which is measured by the number of participant nucleons in the collision process. So, in an effort to eliminate the trivial size dependence, a second measure of the skewness can be obtained.
	\begin{align}
	\Gamma_{p_{\rm T}}=\frac{\left\langle \Delta p_{\rm i}\Delta p_{\rm j} \Delta p_{\rm k}\right\rangle\left\langle \! \left\langle p_{\rm T}\right\rangle \!\right\rangle}{{\left\langle \Delta p_{\rm i}\Delta p_{\rm j}\right\rangle }^2}
	\label{intenskew_eq}
	\end{align}
	$\Gamma_{p_{\rm T}}$ is called intensive skewness and is independent of the  number of participant nucleons or centrality.
	\section{Results and Discussion}
	
	The present analysis has been carried out with 100 million events, generated for p$-$p collisions at  $\sqrt{s} = $ 7 TeV and 13 TeV using PYTHIA 8 event generator with the standard Monash 2013 tune. The charged particles were accepted in a pseudo-rapidity window of $|\eta| < 0.8$. Further, charged particles having transverse momentum range $0.15<p_{\rm T}<2.0 $ GeV/\textit{c} were considered for this investigation. The event multiplicity estimation was done by classifying the events based on the charged particle multiplicity obtained in the forward (backward) pseudo-rapidity ranges : $2.8<\eta<5.1$ and $-3.7<\eta<-1.7$. The obtained multiplicity values scale linearly with  the charged particle multiplicity in the accepted range ( $|\eta| < 0.8$), used for the analysis. For each event class, the corresponding mean charged multiplicity, $\langle N_{\rm ch} \rangle$ was obtained in $|\eta| < 0.8$.  This was done to avoid auto-correlation biases.
	\begin{figure}[h!]
		\centering
		\includegraphics[width=0.9\linewidth]{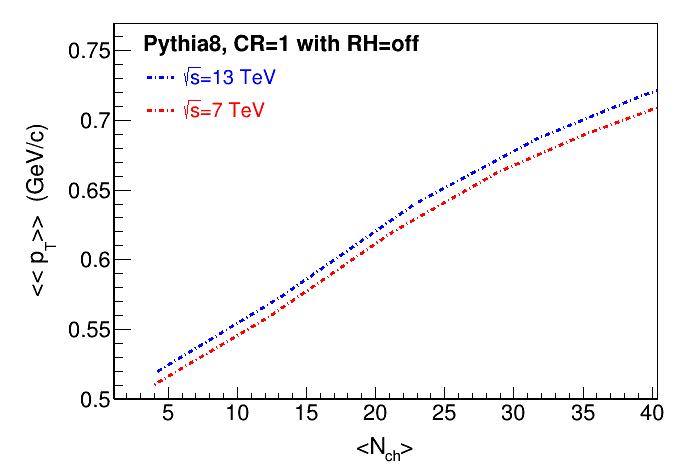}
		\includegraphics[width=0.9\linewidth]{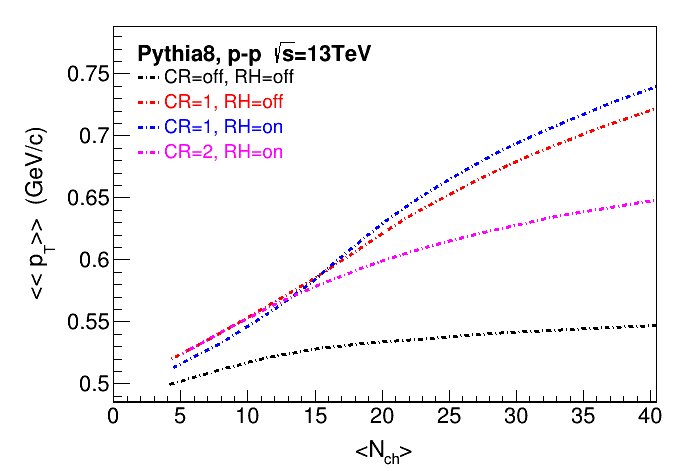}
		\includegraphics[width=0.9\linewidth]{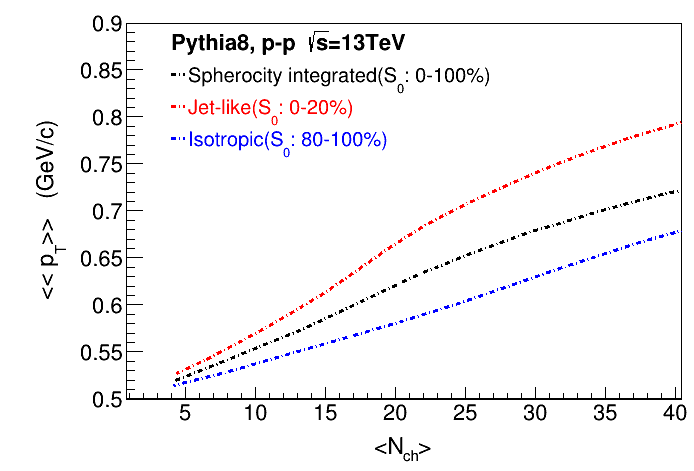}
		\caption{The top panel shows the variation of $\langle\!\langle p_{\rm T} \rangle\!\rangle$ with $ \langle N_{\rm ch} \rangle$ for p$-$p collisions at  $\sqrt{s} = $ 7 TeV and 13 TeV.  The effect of different modes of  color reconnection (CR) on the variation of $\langle\!\langle p_{\rm T} \rangle\!\rangle$ with multiplicity for  collision energy $\sqrt{s} = $ 13 TeV is shown in the middle panel. The bottom panel shows  the variation 
			of $\langle\!\langle p_{\rm T} \rangle\!\rangle$ with $ \langle N_{\rm ch} \rangle$ for p$-$p collisions at $\sqrt{s} = $ 13 TeV for different spherocity classes.}
		\label{meanpt}
	\end{figure}
	\par
	Figure \ref{meanpt} shows the variation of $\langle \! \langle p_{\rm T} \rangle\! \rangle$  as a function of  mean charged particle multiplicity $\langle N_{ch}\rangle $ within $|\eta|<0.8 $ for p$-$p collisions at $\sqrt{s}= $ 7 TeV and 13 TeV. The effect of different modes of color reconnections has been shown in the middle  panel of Figure \ref{meanpt} for  13 TeV beam energy. One can observe that the $ \langle\!\langle p_{\rm T}\rangle\! \rangle$  values are systematically higher for higher collison energy. It can also be observed that switching on the color reconnection (CR) modes increases the $\langle\! \langle p_{\rm T} \rangle\!\rangle$ value of the charged particles. The implementation of CR models in PYTHIA 8 was found to mimic several collective-like behaviour seen in heavy-ion collisons and thus, one is prompted to study the effects of different CR models on the fluctuation observables. The mean transverse momentum, $\langle \! \langle p_{\rm T} \rangle\!\rangle$ of the charged particles shows a clear rise with the average charged particle multiplicity ($\langle N_{\rm ch}\rangle $), for all CR modes, compared to the case when colour reconnection is turned off (middle panel). The QCD-based CR model (CR=1) with the rope formation exhibits the strongest upward trend.  It should be mentioned that the mechanism of color reconnection was originally introduced to explain the  increase of $\langle \! \langle p_{\rm T} \rangle\!\rangle$ with charged particle multiplicity. It allows the partonic  interactions through color strings from different semi-hard scatterings. There is a prior fusion of strings from different
	multi-partonic systems before hadronization, which leads to a decrease in overall multiplicity but the particles are more energetic and consequently have a higher value of $\langle \! \langle p_{\rm T} \rangle\!\rangle$.
	\par 
	The bottom plot in the Figure \ref {meanpt} shows the variation of $\langle\! \langle p_{\rm T} \rangle\!\rangle$  as a function of average charged particle multiplicity for different spherocity classes.  One can observe that  the values are significantly higher for the events representing the lower 0-20\% spherocity class (Jetty). These are the events dominated by hard scatterings and give rise to di-jet topology. The values for isotropic events (80-100\% spherocity class) are consistently lower than those for spherocity integrated (0-100\% spherocity) event class. The behaviour is qualitatively similar to the one measured by ALICE experiment \cite{mptalice}. This can be attributed to the dominance of hard scattering events in the lower 0-20\% spherocity class, while the 80-100\% are dominated by underlying events.
	\par
	As mentioned earlier, the event-by-event mean transverse momentum fluctuations can be quantified using two-, three- and four- particle correlators. These fluctuations  mainly arises from the initial state fluctuations and 
	contain relevant information about the underlying physical processes of particle production \cite{skewness}.
	\begin{figure}[h]
		\centering
		\includegraphics[width=0.9\linewidth]{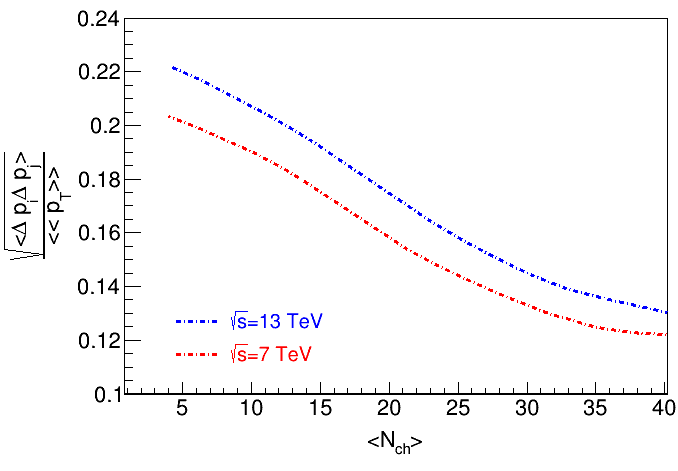}
		\includegraphics[width=0.9\linewidth]{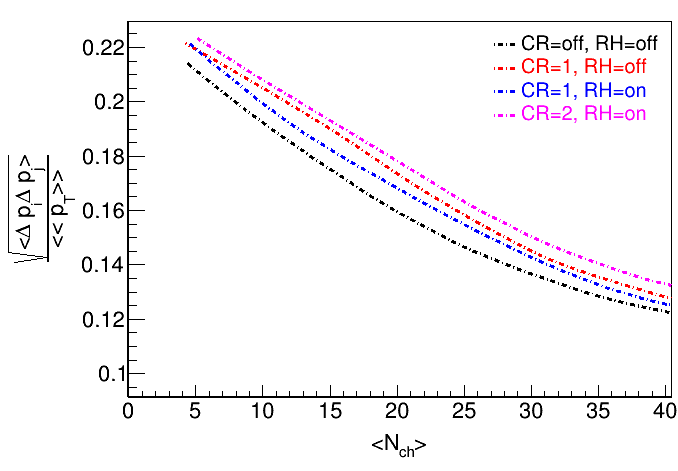}
		\includegraphics[width=0.9\linewidth]{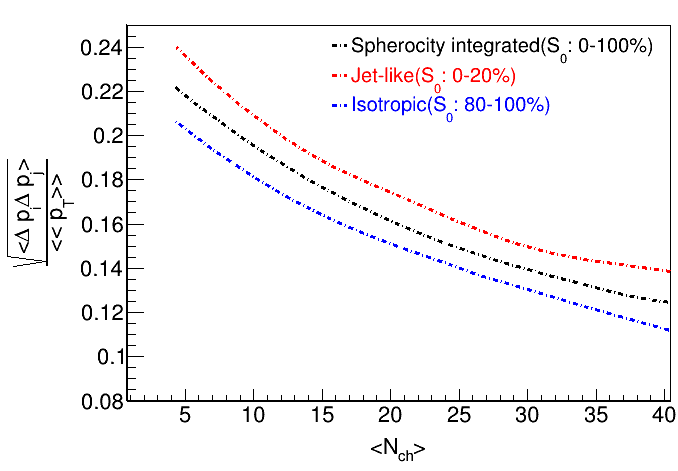}
		\label{twopart}
		\caption{ The variation of two-particle correlator at $\sqrt{s} = $ 7 and 13 TeV as a function of $ \langle N_{\rm ch} \rangle$. The middle panel shows the effect of different color reconnection models at $\sqrt{s} = $13 TeV. The bottom panel shows the variation of two-particle correlator with $ \langle N_{\rm ch} \rangle$ in different spherocity classes.}
		\label{twopart}
	\end{figure}
	\par
	Figure \ref{twopart} depicts the fluctuation of the two-particle correlator for p$-$p collisions at $\sqrt{s} = $ 7 and 13 TeV as a function of charged particle multiplicity. It can be observed that the strength of the two-particle correlator decreases with an increase of the average multiplicity. This can be attributed to the independent superposition of multi-partonic interactions in p$-$p collisions. However, the values at $\sqrt{s} = $13 TeV are systematically higher than that observed at 7 TeV, indicating an energy dependence of the observable. The middle panel of the Figure \ref{twopart} shows the effect of various modes of color reconnections on the strength of the correlator. The mechanism of color reconnection increases the correlation strength due to enhanced partonic interactions via color strings. 
	The interplay of rope formation in different CR modes  also causes a difference in the strength of the correlator.   
	The variation of the two-particle correlator  as a function of multiplicity was also compared for three different spherocity classes as shown in Figure  \ref{twopart}. One can observe that the values are higher for the lower 0-20\% spherocity class (Jetty), indicating that significant contribution comes from the di-jets in low spherocity classes. \par
\begin{figure}[!h]
		\centering
		\includegraphics[width=0.9\linewidth]{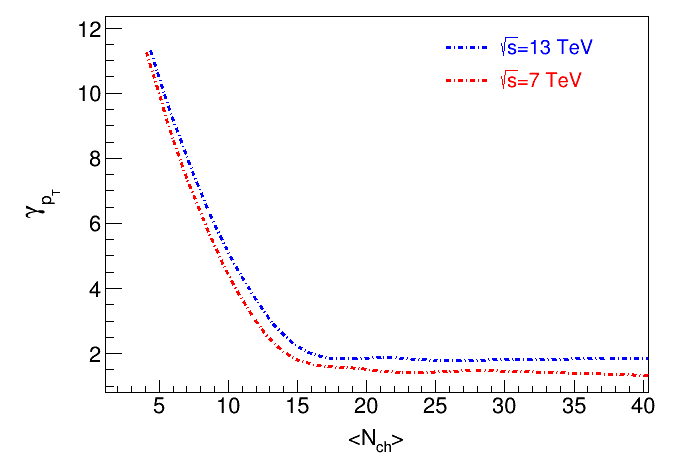}
		\includegraphics[width=0.9\linewidth]{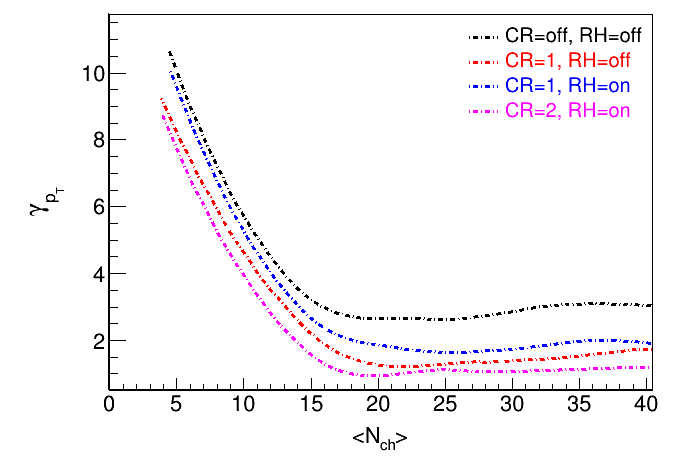}
		\includegraphics[width=0.9\linewidth]{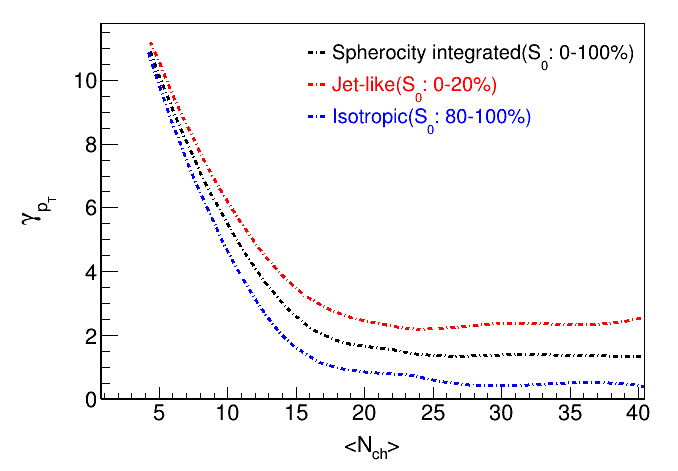}
		\caption{ The energy dependence (top) and the effect of different color reconnection (CR) model comparison (middle) of $\gamma_{p_{\rm T}}$ in p$-$p collisions at  $\sqrt{s} = $ 7 TeV and 13 TeV. The bottom panel shows its variation in different spherocity classes at $\sqrt{s} = $ 13 TeV.}
		\label{stand_skew}
	\end{figure}
\begin{figure}[!h]
		\centering
		\includegraphics[width=0.91\linewidth]{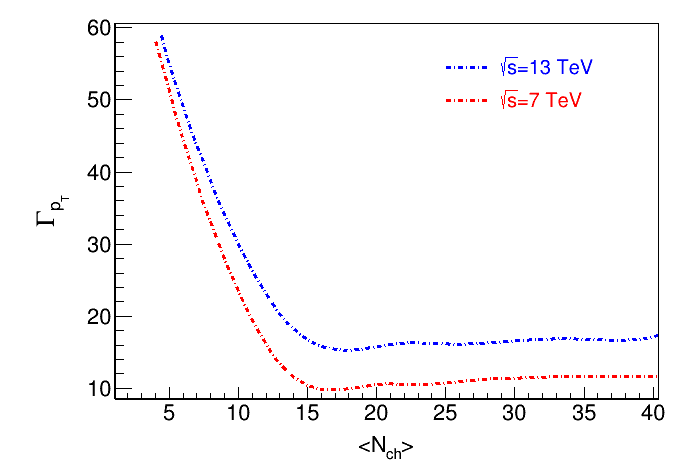}
		\includegraphics[width=0.9\linewidth]{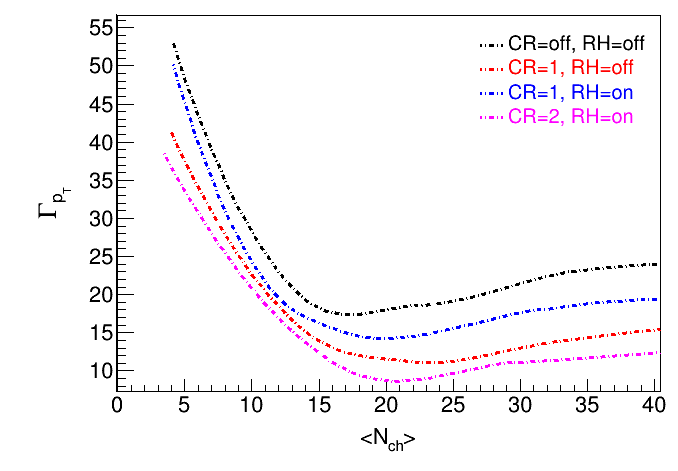}
		\includegraphics[width=0.9\linewidth]{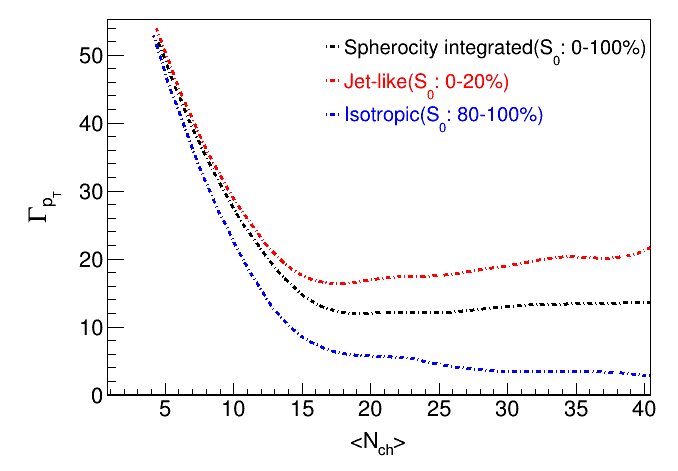}
		\caption{ The energy dependence (top) and the effect of different color reconnection (CR) model comparison (middle) of $\Gamma_{p_{\rm T}}$ in p$-$p collisions at $\sqrt{s} = $ 7 TeV and 13 TeV. The bottom panel shows its variation in different spherocity classes at $\sqrt{s} = $13 TeV.}
		\label{inten_skew}
	\end{figure}

	Recently, it was shown in the reference \cite{skewness} that event-by-event fluctuations of the $\langle\!\langle p_{\rm T}\rangle\!\rangle$ in heavy-ion collisions exhibit a positive skew, as predicted by hydrodynamic evolution of the medium. The idea stemmed from the strong correlation of transverse momentum fluctuations with the fluctuations of the initial energy of the fluid. 
	An estimation of  the three-particle correlation using standardized and intensive skewness for p$-$p collisions at $\sqrt{s} = 7 $ and  13 TeV  is shown in the Figure  \ref{stand_skew} and \ref{inten_skew}, respectively.
	Assuming that the dynamical and statistical fluctuations scale inversely  with multiplicity \cite{skewness},  the standardized skewness is expected to be larger for lower multiplicity classes. It is indeed observed that $\gamma_{p_{\rm T}}$ exhibits a sharp diminishing trend upto a charged particle multiplicity range of 15 and saturates thereafter.  However, the magnitude is considerably higher when compared to hydrodynamic predictions for heavy-ion collisions. This trend is also seen for intensive skewness, $\Gamma_{p_{\rm T}}$ which was introduced to eliminate the volume dependence in heavy-ion collisions. Although, p$-$p system is not affected by volume fluctuations, studying the variation of $\Gamma_{p_{\rm T}}$ is interesting to form a baseline. Like $\gamma_{p_{\rm T}}$, the values of $\Gamma_{p_{\rm T}}$ are higher for low multiplicity events and gradually saturates after 15. In addition, the comparisons of various CR models (middle panels of Fig. \ref{stand_skew}, \ref{inten_skew}) revealed a decrease in the correlation strength when CR was introduced.  This could indicate that  the color string fusion dominantly affects (or enhances) the two-particle correlation and does not have any significant effect on higher order correlations. The spherocity comparisons of the three-particle correlator indicates  higher correlation for jetty events (0-20\% spherocity class), whilst the isotropic events (80-100\% spherocity class) have lower values. (Fig. \ref{stand_skew} and \ref{inten_skew}).
	\begin{figure}[!h]
		\centering
		\includegraphics[width=0.9\linewidth]{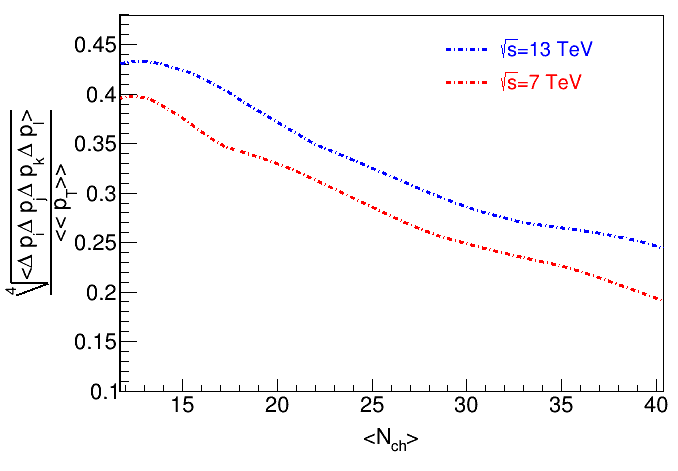}
		\includegraphics[width=0.9\linewidth]{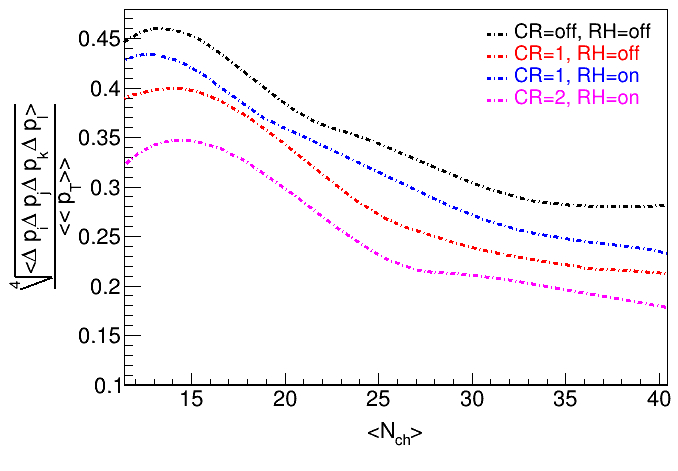}
		\includegraphics[width=0.9\linewidth]{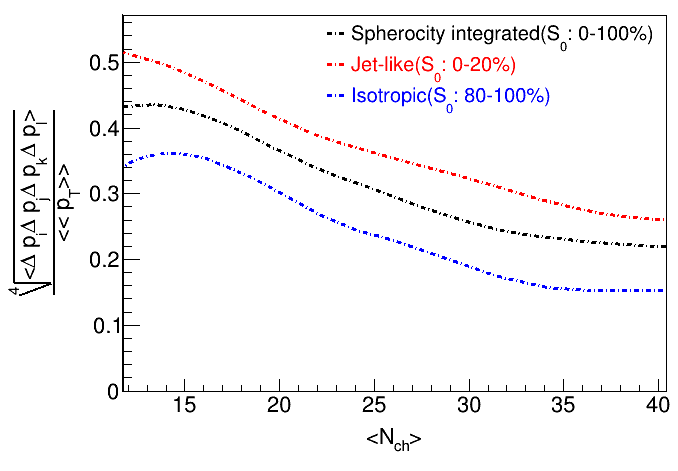}
		\caption{ The four-particle correlator as a function of $\langle N_{\rm ch} \rangle$ for different energies (top) and color reconnection (CR) models (middle). The bottom panel shows the correlator strength in different spheocity classes at $\sqrt{s}=13$ TeV.}
		\label{fourpart}
	\end{figure}
	\par
    The study was further extended to observe the variation of the four-particle correlator with mean charged particle multiplicity for p$-$p collisions at $\sqrt{s}= 7$ and 13 TeV as shown in Figure \ref{fourpart}. These types of studies are subjected to large statistics and can reveal crucial information on the higher order correlation strength. It is observed that the correlator shows a smooth decreasing trend towards higher multiplicity classes. The increase in parton activity that causes the parton's terminal strings to recombine$-$ known as color reconnection, produces a weaker trend in the correlation function. This trend is weakest in case of a color reconnection model CR=2 which is based on the movement of gluons in order to produce a string of minimum tension. Further, spherocity investigation shows that the correlations are strongest for events dominated by hard scattering (lower 0-20\% spherocity class). 
    
    \section{Summary}
    Event-by-event fluctuation of the mean transverse momentum ($\langle p_{\rm T} \rangle$) has been studied as a function of average charged particle multiplicity ($\langle N_{\rm ch} \rangle$) in p$-$p collisions at $\sqrt{s}$ = 7 TeV and 13 TeV  using the PYTHIA 8 event generator. The charged particles were selected from a kinematic range of $0.15<p_{\rm T}<2$  GeV/\textit{c} and $|\eta| < 0.8$. The dynamical fluctuations of the mean transverse momentum is caused by fluctuation in the early stages of the collision systems and have been measured using two-, three-, and four-particle correlators. A characteristic decrease of the two-particle particle correlator with the charged particle multiplicity is observed. This phenomenon is attributed to independent superposition of multi-partonic interaction in p$-$p collisions. The effects of various microscopic processes like color reconnection and multi-partonic interactions has also been investigated. The mechanism of color reconnections with rope formation scenario showed higher correlation. Higher order (third and fourth) correlation functions, which have been reported for the first time, did not, however, exhibit the same behaviour.
    The study also suggests an energy dependence of the mean transverse momentum fluctuation in p$-$p collisions at $\sqrt{s}$ = 7 TeV and 13 TeV. Furthermore, the  fluctuation observable has also been studied in different intervals of transverse spherocity to understand the relative contributions originating from hard scattering and underlying events. It was observed that  the correlations were always stronger for events with di-jet-like topology. The present 
    study in p$-$p collisions would act as a baseline for the upcoming measurements in A$-$A as well as p$-$p collisions at the LHC.
    \section{Acknowledgements}
    The authors would like to thank the Department of Science and Technology (DST), India for supporting the present work.


\begin{thebibliography}{50}
    	
    	\medskip
    	
    	\bibitem{star2005} J. ~Adams et~al, STAR Collaboration,  Phys. Rev. {\bf C  72}, 044902 (2005).
    	\bibitem{alice2014} B.~B.~Abelev et~al, ALICE Collaboration , Eur. Phys. J.  {\bf C 74}, 3077 (2014).
    	\bibitem{trainor1} T. ~A. Trainor,  Phys. Rev. {\bf C  92}, 024915 (2015).
    	\bibitem{stodolsky} L. ~Stodolsky,  Phys. Rev. Lett. {\bf 75}, 1044 (1995).
    	\bibitem{star2019} J. ~Adams er~al, STAR Collaborationl,  Phys. Rev. {\bf C  99}, 044918 (2019).
    	
    	\bibitem{mptvoloshin} S.~Voloshin, Phys. Lett.  {\bf B 632}, 490-494 (2006).
    	\bibitem{schenke} B.~Schenke, Chun Shen and Derek Teaney, Phys. Rev. {\bf C  102}, 034905 (2020).
    	
    	
    	\bibitem{mpi}P. Bartalini, E. Berger, B. Blok, G. Calucci, R. Corke, et al., arXiv:1111.0469 [hep-ph]
    	\bibitem{pythia8} Torbj{\"o}rn Sj{\"o}strand, Stefan Ask, Jesper~R Christiansen, Richard Corke, Nishita Desai, Philip Ilten, Stephen Mrenna, Stefan Prestel, Christine~O Rasmussen, and Peter~Z Skands, Comput. phys. commun. {\bf 191}, 159--177 (2015).
    	
    	
    	
    	\bibitem{mptalice} B.~B.~Abelev et al, ALICE Collaboration, Phys. Lett.  {\bf B 727}, 371 (2013).
    	
    	
    	\bibitem{color0} Torbj{\"o}rn Sj{\"o}strand, arXiv:1310.8073,(2013). 
    	\bibitem{color1} Christian ~Bierlich and Jesper Roy Christiansen, Phys. Rev {\bf D  92},094010 (2015). 
    	\bibitem{color2} Jesper Roy Christiansen and P.~Z.Skands , JHEP, {\bf 08}, 003 (2015). 
    	\bibitem{rope1} T. ~S.~Biro,~H.~B.~Nielson and ~J.~Knoll, Nucl. Phys. {\bf B 245}, 449, (1984).
    	\bibitem{rope2} C.Bierlich, G. Gustafson, L. Lonnblad and A. Tarasov, J. High Energ. Phys. {\bf 2015: 148}, (2015).
    	\bibitem{sdashrope1} Ranjit Nayak, Subhadip Pal, and Sadhana Dash, Phys. Rev. {\bf D 100}, 074023 (2019).
    	\bibitem{sdashrope2} Pritam Chakraborty  and Sadhana Dash, Phys. Rev. {\bf C 102}, 055202 (2020).
    	\bibitem{sdashrope3} Ankita Goswami, Ranjit Nayak, Basanta Kumar Nandi,  and Sadhana Dash, Eur. Phys. J. {\bf C 81}, 988 (2021).
    	\bibitem{epos} T. Pierog, Iu. Karpenko, J. M. Katzy, E. Yatsenko, and K. Werner, Phys. Rev  {\bf C  92}, 034906 (2015).
    	
    	
    	\bibitem{sphero1} A. Banfi, G. P. Salam, and G. Zanderighi, JHEP {\bf  06}, 038 (2010).
    	\bibitem{sphero2} A. Ortiz, G. Paic, and E. Cuautle,  Nucl. Phys. {\bf A 941} , 78-86 (2015).
    	\bibitem{sphero3} S. ~Acharya et~al, ALICE Collaboration , Eur. Phys. J.  {\bf C 79}, 857 (2019). 
    	\bibitem{skewness} Giuliano Giacalone, Fernando G. Gardim, Jacquelyn Noronha-Hostler, and Jean-Yves Ollitrault, Phys. Rev. {\bf C  103}, 024910 (2021).
    	
    	
    	
    	
    	
    	
    \end{thebibliography}
\end{document}